# Qualitative and quantitative evaluation of COVID-19 outbreak severity with the use of meta-projections based on Richards' curve parameters

Evagoras Xydas[1], Konstantinos Kostas[2]

[1]*IREROBOT LTD, Nicosia, Cyprus*
[2]*Nazarbayev University, Nur-Sultan Kazakhstan*

## Abstract

By April 2020, the first wave of COVID-19 outbreak has reached a plateau in some countries, while in other countries it is still increasing exponentially with varying doubling rates, forcing governments to continuously evaluate the situation and adapt the implemented control measures accordingly. Researchers have shown that even simple empirical models stemming from biological growth modeling have the potential to provide useful information on the development and severity of ongoing epidemics since they can be employed as tools for carrying out projections on the size of the affected population, timing of turning points, as well as best- and worst-case scenarios. Nevertheless, they commonly exhibit considerable sensitivity to some input parameters' variance which results in large fluctuations in the generated projections, thus rendering predictions difficult and even risky. In this work we examine a novel meta-projections-based approach which allows us to evaluate the model's current trends and assess whether generated projections are at a transient or stable state. Meta-projections can be extracted from graphs of successive estimations of model's parameters and resulting projections, over a sequence of days being gradually added to the employed model. In other words, projections are carried out on truncated timeseries of cumulative numbers of confirmed cases with increased lengths at each successive evaluation. This allows us to trace the values of model parameters over a certain period of time and examine their trends which may converge to specific values for well-described, settled-growth cases or exhibit a changing or even an erratic behavior for cases that undergo epidemiological transitions and/or are inappropriately described by the current model instance(s). We have computed meta-projections and compared our findings for countries at different stages of the epidemic with stable or unstable behaviors and increasing or decreasing numbers of confirmed cases. Our results indicate that meta-projections can aid researchers in assessing the appropriateness of their relevant models and in effect decrease the uncertainty in their estimations of an epidemic's severity and development.

# Introduction

By April 2020, the first wave of COVID-19 outbreak has reached a plateau in some countries, while it is still in a transient phase with varying doubling rates, in others [1],[2]. Countries that have managed to control the outbreak are working on avoiding a second outbreak while countries with ongoing pandemic development are striving to contain the disease dispersion in the society. Empirical growth models such as Richards' [3] (aka the Generalized Logistic Curve) have shown the potential to evaluate the severity of outbreaks. In retrospective studies, Richards' model has been demonstrated to fit well to epidemic data, such as those coming from the Severe Acute Respiratory Syndrome (SARS) outbreak in Taiwan [4]. Zhou and Yan used the Richards' curve to model the confirmed cumulative number of SARS cases in Singapore, Hong Kong, and Beijing [5]. In another work, Hsieh and Cheng [6] employed the Richards' model to fit the SARS data for Toronto where a multi-stage outbreak was recorded. It was established that the model can be effectively employed to project turning points and to demonstrate the disease severity in multi-wave epidemics. In another work, Hsieh et al. [7] employed the Richards' model to fit daily cumulative cases' data from the 2003 SARS outbreaks in Taiwan, Beijing, Hong Kong, Toronto, and Singapore. Using different lengths of the respective time series, they demonstrated that the model can be used to analyze the epidemic retrospectively and identify the significance of various events occurring at different times during the outbreak, and further, that the carrying capacity, $K$ (or the maximum projected number of cumulative cases), can be estimated in real-time by the method.

In a more recent work, Roosa et al. [8] used phenomenological models including the Richards' model to generate and evaluate short-term forecasts of the cumulative number of confirmed cases in Hubei province, the epicenter of COVID-19 in China. Their findings suggest that the containment strategies implemented in China have successfully reduced transmission, and in addition, they were able to forecast the slowing down of confirmed cases.

In general, previous works have established the ability of Richards' model to evaluate the outbreak severity and to forecast the turning points as well as the maximum number of cases during an epidemic. Essentially it is a statistical fit with saturation in growth which takes place as the susceptible population decreases, control measures (such as social distancing) are implemented, etc. Thus, the Richards' model encompasses the consequences of the mechanisms of epidemic evolution by explicitly incorporating response mechanisms in the equations. On the other end, the SIR (Susceptible, Infected and Removed) model considers the mechanisms of transmission within the population. One of differences between the SIR and Richards' model is that the employed exponent in the latter does not seem to have a direct epidemiological correspondence, in contrast to SIR model where all of its parameters exhibit direct physical interpretations. It has been shown, however, that the exponent of Richards' model does have an indirect epidemiological explanation with reference to the SIR model [10]. In this work we built on previous studies and propose an approach by which, in addition to real-time forecasts and projections, successive estimations of the model parameters and analysis of their history allow us to identify phases of stability or epidemiological transitions in the modeled case. Additionally, converging parameter histories allows, up to a certain degree, to establish the reliability of the current forecasts. Our results are compared to the SIR model [12], using an implementation by Joshua McGee [11], [12], and they are further analyzed with reference to the implemented COVID-19 containment measures in each country.

## Background

### *Generalized Logistic Function (Richards' curve)*

The Richards' curve or generalized logistic function is an extension of the logistic model proposed by Verhulst in 1838 as a function that can potentially predict population growth [13]. In 1959, Richards examined three best-known 'growth functions' at the time, i.e., the monomolecular, autocatalytic and Gompertz, and proposed a generalized form that included all of the above while allowing meaningful interpretations for the employed constants / parameters [14]. The generalized logistic function, expressing the value of a growing quantity *I* at time *t*, can be written as

$$I(t) = A + \frac{K - A}{(C + Qe^{-B(t-t_M)})^{1/v}},\qquad(1)$$

where *A*, *K* are the lower and upper[1] asymptotes, respectively, *B* the growth rate, *v* affects the position of the maximum growth along the curve and needs to assume positive values, *C* typically assumes the value 1 so that *K* equals the upper asymptote[2], *Q* relates to the value of *I* at t=0, and $t_M$ corresponds to the time at which *I* equals to $A + \frac{K-A}{(C+1)^{1/v}}$.

A special case of the generalized logistic function (with *A*=0 and *C*=1) is

$$I(t) = \frac{K}{(1 + Qe^{-rv(t-t_0)})^{1/v}}\qquad(2)$$

that corresponds to the solution of the so-called Richards' differential equation:

---

[1] Assuming C equals to 1.
[2] If C has a value other than 1, the upper asymptote will be equal to $A + \frac{K-A}{\sqrt[v]{C}}$

$$I'(t) = rI\left(1 - \left(\frac{I}{K}\right)^v\right) \tag{3}$$

with an initial condition of $I(t_0)=I_0$. This differential equation coincides with the logistic model's differential equation, if $v$ equals 1 and it is exactly this addition of parameter $v$ that enables the additional flexibility in Richards' model. Finally, assuming that $t_0=t_M$ and given that $Q=-1+(K/I_0)^v$, it is easy to derive the following equivalent expression from Eq.(2):

$$I(t) = \frac{K}{(1 + e^{-B(t-t_M)})^{1/v}}, \tag{4}$$

where $B=rv$. Furthermore, it is also straightforward to see that the inflection point, or point of maximum growth, for Eq.(4) is equal to $t_M-log(v)/B$, which further simplifies to just $t_M$ when $v=1$.

Figure 1 depicts a series of sample curves generated using Richards' model. The resulting curves exhibit a sigmoid shape with two asymptotes and an inflection point which in the case of our epidemiological data represents the turning point, i.e., the point with the maximum number of daily cases.

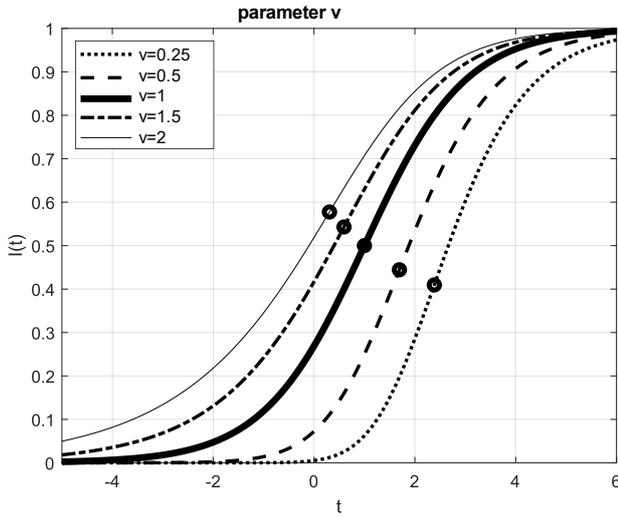
(a) Sigmoid curves generated by Eq.(4) with $K=B=t_M=1$ and varying values of $v$.

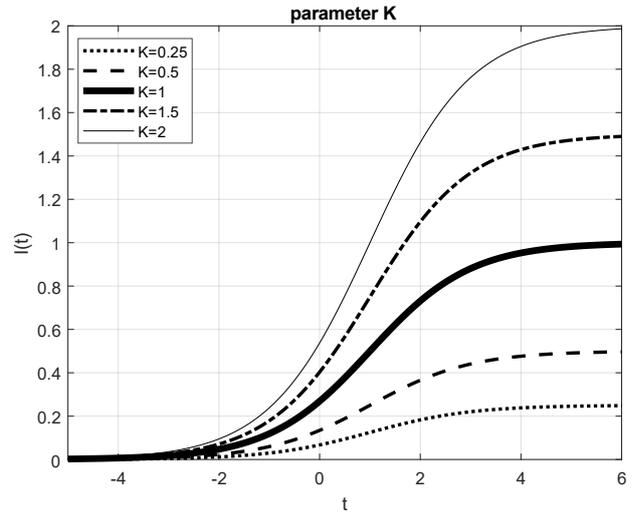
(b) Sigmoid curves generated by Eq.(4) with $B=t_M=v=1$ and varying values of $K$.

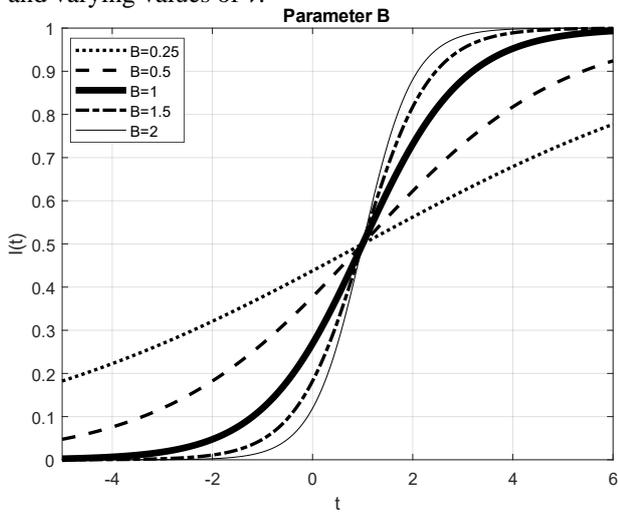

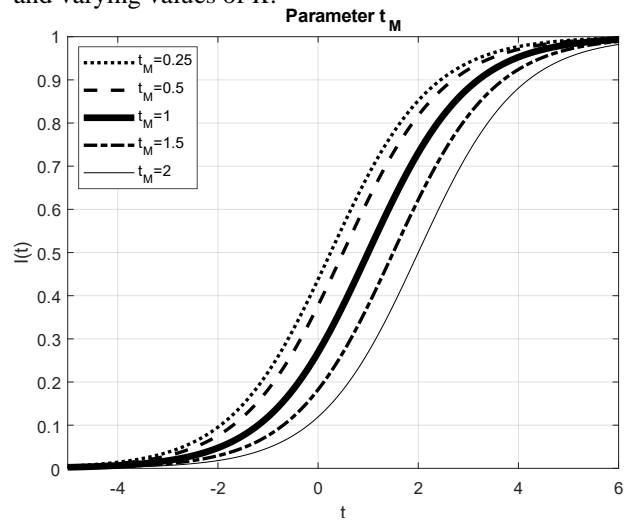

**(c)** Sigmoid curves generated by Eq.(4) with *K=t_M=v=1* and varying values of *B*.

**(d)** Sigmoid curves generated by Eq.(4) with *K=t_M=v=1* and varying values of *B*.

**Figure 1**: Sigmoid curves generated by Eq.(4). Notice that $t_M$ corresponds to inflection point's abscissa when *v=1*. For values of *v* less than 1 the inflection point moves towards the lower asymptote while for *v* greater than 1 the inflection point approaches the upper asymptote.

For the remaining of this document we will use the term *Richards' curve* and *Richards' model* to refer to the family of curves generated by the specific expression appearing in Eq.(4).

## *SIR Model*

The SIR model is one of the simplest compartmental models and it is the basis for many of its derivatives. In SIR, we consider three groups of people: the susceptible (S), which are the people who can get infected, the infected (I), which have already contracted the disease, and the removed (R) that represent the recovered and diseased people. The model assumes that no one in the removed category will contract the disease. Also, it assumes that within the outbreak period, no significance population change takes place (e.g., through new births, deaths, migration etc.). The model is described by the following three non-linear differential equations, describing the change in the categories with respect to time:

$$\frac{dS}{dt} = -bSI \tag{5}$$

$$\frac{dI}{dt} = bSI - aI \tag{6}$$

$$\frac{dR}{dt} = aI, \tag{7}$$

where, a and b are positive constants. More specifically, *b*, the infection rate, is the average rate by which an infected individual transmits the disease to susceptible individuals and *a* is the recovery rate. The recovery rate is equal to the reciprocal over the duration of the illness. Since, we have assumed that our population N(t)=1 does not change over time, it is obvious that:

$$S(t) + I(t) + R(t) = N(t) = N = 1\ (100\%), \tag{8}$$

where *N* is the assumed constant population size and obviously $\frac{dS}{dt} + \frac{dI}{dt} + \frac{dR}{dt} = 0$. Note that the population size does not necessarily correspond to the whole population, e.g., of a country, but rather the part of population affected by the disease. The basic reproduction rate ($R_0$) is defined to be:

$$R_0 = \frac{b}{a}\frac{S(0)}{N} \approx \frac{b}{a}. \qquad (9)$$

At the beginning of the pandemic $\frac{S(0)}{N}$ is equal or approximately equal to 1 since there might be a small number of infected people already. The role of the reproduction rate is extremely important as for values of $R_0$ higher than 1, an epidemic outbreak occurs, while when $R_0$ is less than one, the disease will phase out regardless of the initial size of the susceptible population. The effective reproduction number represents values that change over time as the susceptible population changes. If the effective reproduction number becomes less than 1, then the disease will die out. Essentially, the basic reproduction number is the maximum value of the effective reproduction number which is calculated as an infected individual enters a fully susceptible population. It is clear that in contrast to the epidemiological growth models, the SIR model takes account the transmission mechanisms. Given the success of Richards' model to accurately fit previous epidemics, it is no surprise that direct non-linear relations were established between its parameters and the parameters of the SIR model as well as the reproduction number [10]. As mentioned above, in this work, we use an implementation [11] of the SIR model presented in 24[12].

## Methodological Approach

### Fitting Richards' curve to epidemiological data

One of the most straightforward ways for fitting Richards' model to existing epidemiological data, i.e., time-series of cumulative numbers of daily cases $\{(t_i,c_i)\}$, i=1,…,N, is to setup an optimization problem, using for example a least-squares approach, and minimize the deviation of the generated curve from the given dataset. The minimizer $\mathbf{q}^* \in Q \subset R^4$ will then obviously be the solution of the problem and its components will correspond to the parameters (*K, B, $t_m$* and *v*) in Eq.**(4)**. In other words, we need to solve the following minimization problem:

$$\min_{\mathbf{q}} Z(\mathbf{q}) = \min_{\mathbf{q}} \sum_{i=1}^{N}(I(\mathbf{q},t_i) - c_i)^2 \qquad (10)$$

subject to $q_{min} \leq q \leq q_{max}$,

or equivalently $q \in Q \subset R^4, Q = \{\mathbf{q}: q_{min} \leq q \leq q_{max}\}$

Carelessly solving this minimization problem may result in wildly varying **q\*** vectors and corresponding sigmoid curves, which nevertheless assume similar values for the objective function Z(**q\***). Our first line of defense involves an appropriate determination of the lower and upper bounds of **q** appearing as side constraints in Eq.(10). Further to this, and since we employ a sequential quadratic approach for solving the problem, we need to estimate an appropriate starting vector $q_0 \in Q$, which will lead the optimizer efficiently to a reasonable solution. Our overall approach is centered at the determination of $q_0=(K_0, B_0, t_{M0}, v_0)$ as we will subsequently present.

The determination of $q_0$ begins by the simplifying assumption that $v_0$ is equal to 1, i.e., our sigmoid curve is symmetric with an inflection point at $t=t_{M0}$. This assumption allows to directly set the inflection point's estimated abscissa when determining $t_{M0}$. The estimated values for the remaining three parameters $K_0, B_0, t_{M0}$ is determined through the use of a new cases growth diagram (NCG).

In an NCG diagram we are plotting the daily new cases on the y-axis with respect to the total cumulative confirmed cases for the same day plotted on the x-axis; see Figure 2. Both axes use a logarithmic scale and additionally, to avoid daily variability in new cases, daily cases are averaged over the last few days; a period of a week is usually sufficient for smoothing the resulting curve. Obviously, in such a diagram, exponential growth in cases for a particular country, regardless of the exact time-based doubling rate, will appear as a straight line starting from the lower left corner and progressing, with time, towards the upper right corner of the diagram; see e.g., the US or UK trajectories in Figure 2. When the current epidemic dynamics lie in the exponential growth part of the sigmoid curve, commonly used to depict total cases against time, it is very hard to tell whether two-, ten- or hundred-times more confirmed cases will occur at the end. This can be clearly seen if the NCG diagram is compared with the commonly used corresponding depictions in Figure 3. The NCG diagram, as presented here, can provide clear and strong evidence of whether we are in the downturn phase or if we are still riding the "skyrocketing" phase. This diagram is not well-suited for early detection of a downward trend as both the logarithmic scale and the averaging approach distort this, however it provides relatively safe evidence of where the outbreak is heading and this is quite important for our work since we need to be able to come up with a good estimation of the inflection point and when it occurs.

As it is clear from the description above, time is not directly included, but only implied in the NCG diagram. Specifically, time corresponds to the traversal speed of the depicted trajectories or, equivalently, to the density of the points comprising each of them, i.e., a dense point-set corresponds to a slowly growing curve while a sparse one to a fast-growing case. Hence, for determining the time (number of days in our case) at a specific point on the curve, it suffices to count the number of points preceding it[3].

Based on our initial discussion and the NCG diagram, we can now easily estimate the initial parameters' vector and also set the upper and lower boundaries for its components. As mentioned above we begin by assuming that $v_0=1$ with [0.05, 3] being the range of feasible values. Then, for cases where the turning point can be identified on the NCG diagram, i.e., the point with the maximum number of cases before the downturn, we set $t_{M0}$ to be equal to the number of points before the identified maximum, i.e., the number of days between the first case(s) and the maximum number of daily cases.

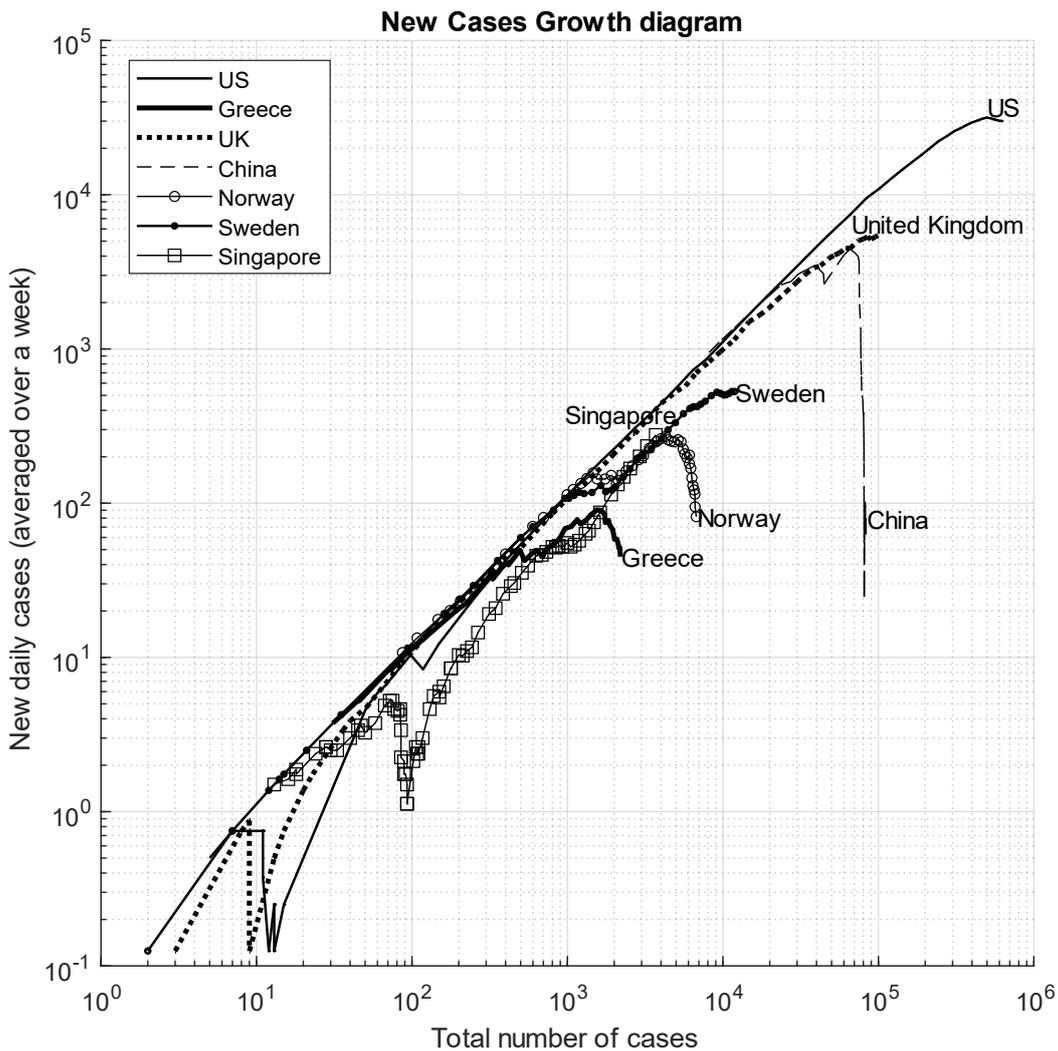

**Figure 2.** New cases growth for several countries on 15.04.2020. China, Norway, and Greece seem to have clearly passed the turning point while US, UK, Sweden, and Singapore are still riding the exponential growth line, although with some first signs that they might be also approaching it.

---

[3] Assuming obviously the same sampling rate for all curves.

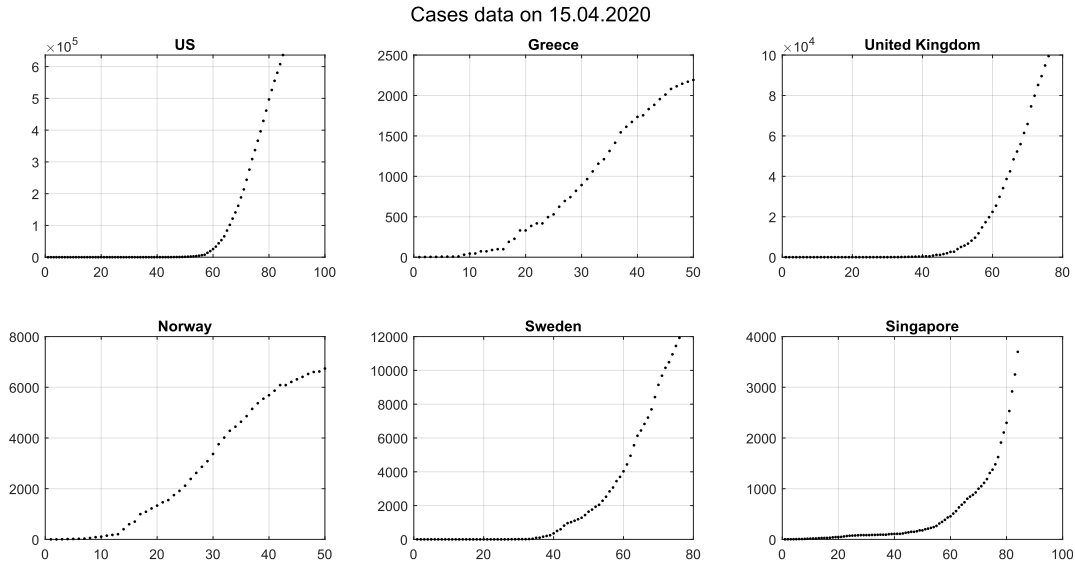

**Figure 3** Plots for total cases against number of days from the first confirmed case. Using these plots to estimate the position of the infection point can be extremely hard. See, for example, the cases of US and UK for which although an indication of reaching the turning point can be seen in the NCG diagram, nothing similar can be easily identified here.

In these cases, the lower and upper limits of $t_M$ can be set few days before and after the estimated day. On the other hand, if the state of the epidemic is still on the exponential growth phase, the lower bound can be set on the last day on the trajectory under examination and the upper bound can lie several days ahead. The next parameter value that needs to be estimated is $K_0$, which, when $t_{M0}$ has been identified and given the fact that $v_0=1$, can be safely assumed to be twice the number of total cases at $t_{M0}$. Obviously, the lower bound can be set as the number of currently reported cases while the upper bound can be estimated as a multiple of the number of cases at the lowest possible position of the inflection point. The final parameter-value and bounds estimation concerns the growth rate $B$. Its initial estimation $B_0$ is computed by averaging the exhibited growth rates on the exponential part of the trajectory while its bounds are determined by the minimum and maximum growth values calculated during the aforementioned averaging process. This set of initial values and bounds estimation produces sigmoid curves with very low root-mean-square error (RMSE), which are in general agreement with the models produced using the SIR approach; see comparisons in the results section.

## Generating meta-projections

Meta-projections can be extracted from graphs of successive estimations of model's parameters and resulting projections, over a sequence of days being gradually added to the employed model. Specifically, for each modeled country we begin our first projection using a data range from the day that the first case occurred till 10 or 20 days after it. The selection of this initial period depends on whether we are dealing with a country that exhibited immediate growth or an initial lag phase. In the successive projections we extend this modelling by one day and solve again the minimization problem in Eq.(10). This iterative process continues till we reach the last day of the time series in question. For each projection we store the optimized vector of parameters ($K$, $t_M$, $B$, $v$), which allows to generate parameter history graphs; see for example Figure 6b.

The analysis of these parameter histories allows us to identify phases of stability or epidemiological transitions in the modeled case. Additionally, converging parameter histories generally increase our confidence in the currently analyzed projection. Further details and discussion of the use of these meta-projections can be found in the following sections.

## Results

The results in this section are generated using data for several countries at various stages of the epidemic. All timeseries have been acquired from the same official and credible sources [1],[2]. Figures 4 and 5 depict the evolution of total cases and deaths numbers, and the deaths over cases ratio, respectively. Figures 6a to 17a depict the projections on 20.04.2020 for each of the selected countries based on both Richards' and SIR models; solid thick and thin line, respectively. Note that the estimated inflection points are depicted on the graphs similarly with vertical thick and thin lines, respectively. Figures 6b to 17b contain depictions of Richards' model parameters history along with the corresponding estimated inflection points, extracted from successive day-by-day projections till April 20[th]. In all of these graphs the horizontal axis refers to time in days, and the vertical axis corresponds to parameter values; see Methods section for more details on the depicted parameters. In the case of $K$, point ordinates represent the projected number of total cases after 120 days from the first day that a confirmed case appeared in each country, which, in most cases, is a good approximation of the upper asymptote. In the inflection-point ($t_i$) graph, point ordinates correspond to each projection's turning-point estimation, in number of days after the first case for each country occurred. Note that the red diagonal line in the same graph is the y=x line and obviously, points above this line place the turning point in the future while points below it, place the estimated inflection point in the past. When these two lines (inflection point estimation history and y=x) intersect, it signifies an estimation that places the inflection point at the same day the projection was made. If the estimations history crosses the red diagonal line on its way down, and stays below it, the turning point is placed in the past and this signifies the transition to the decreasing growth phase. Obviously, if the inflection point is projected in the future, i.e., the curve is above the red line, the epidemic is in its growing phase. Finally, as one can see in Fig. 12b, there can be more than one curve-line intersections in this graph due to changes in transmission dynamics.

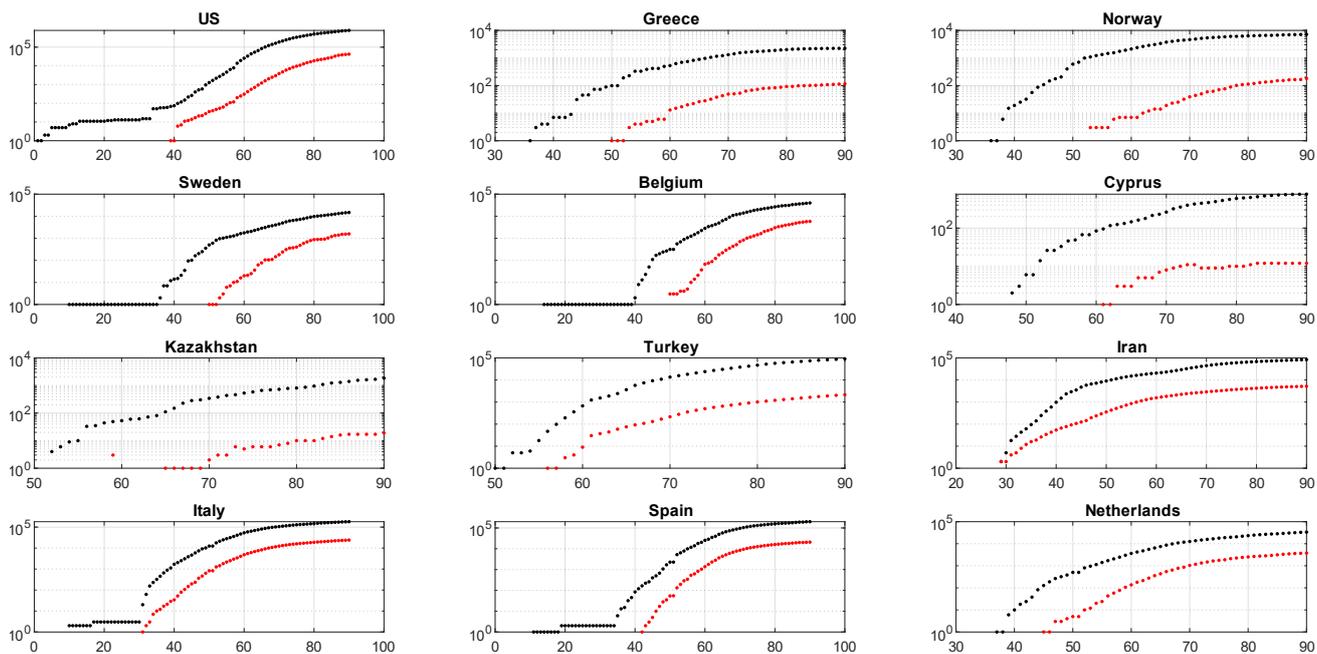

**Figure 4.** Cases and deaths curves for selected countries. Abscissae represent days since the day of the first case while ordinates represent the cumulative numbers of cases and deaths.

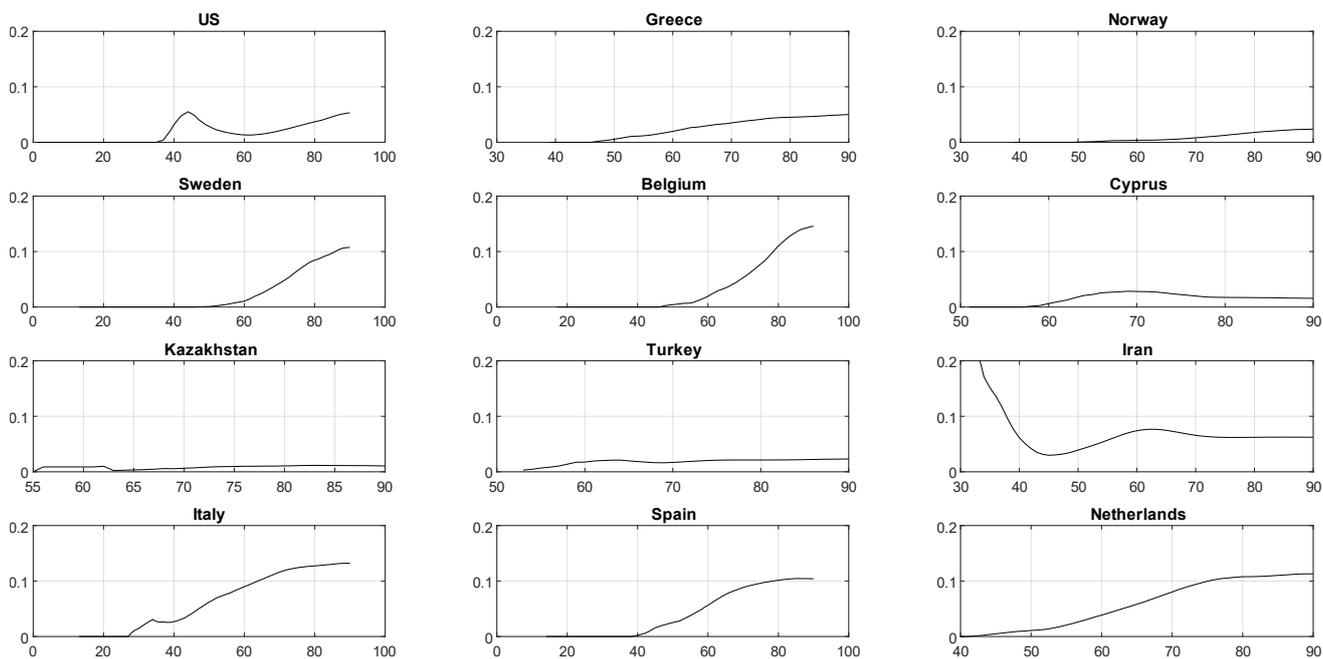

**Figure 5.** Cases and deaths curves for selected countries. Abscissae represent days since the day of the first case and ordinates represent the ratio of deaths over confirmed cases.

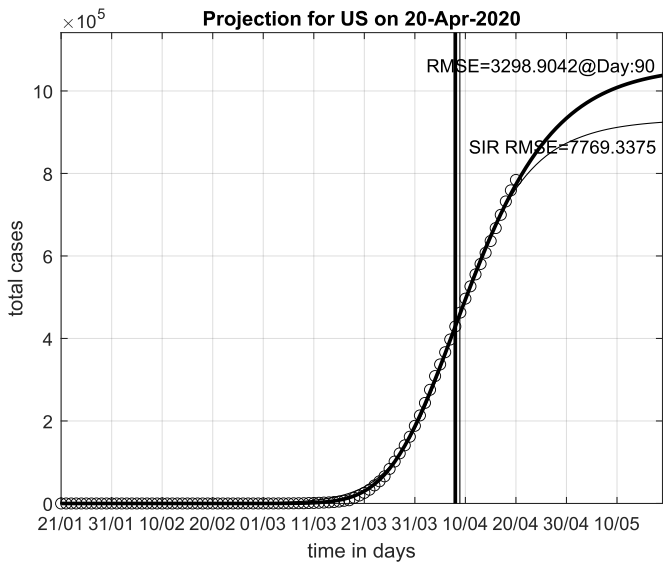

(a) Richards' versus SIR model projections with estimated turning points

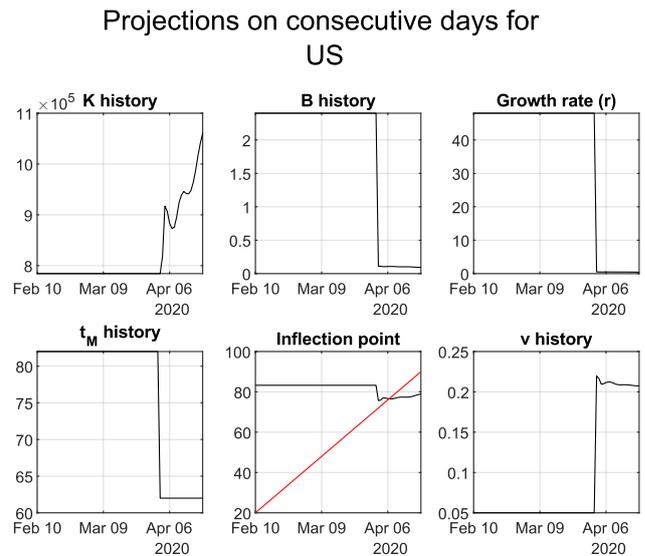

(b) Richards' model parameters history.

**Figure 6.** Modelling results for US on 20.04.2020

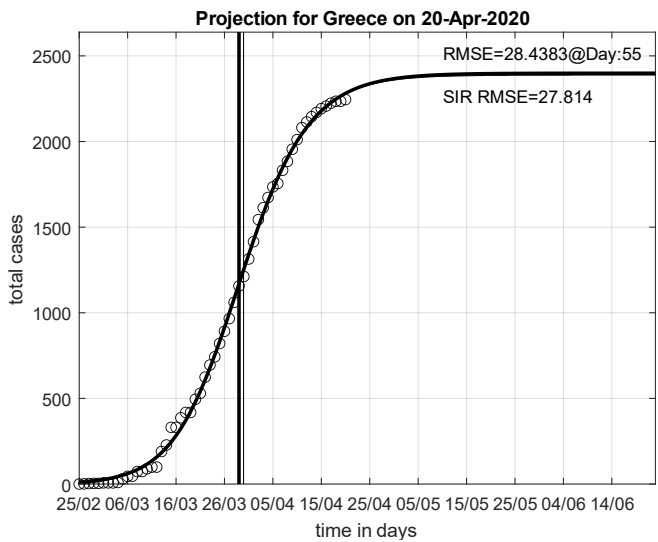

(a) Richards' versus SIR model projections with estimated turning points

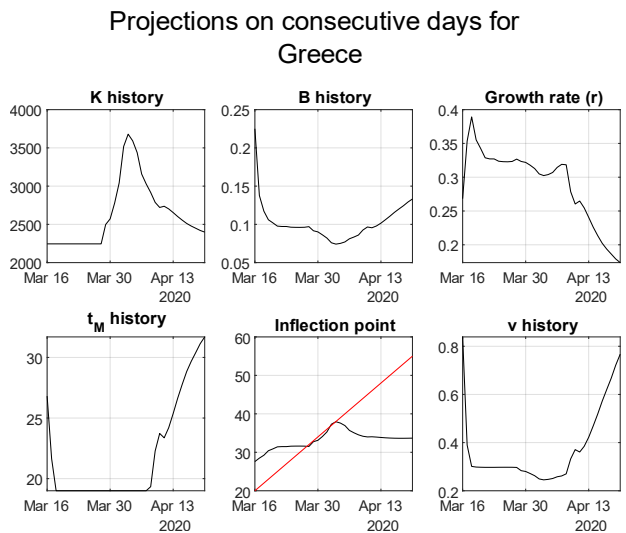

(b) Richards' model parameters history.

**Figure 7.** Modelling results for Greece on 20.04.2020

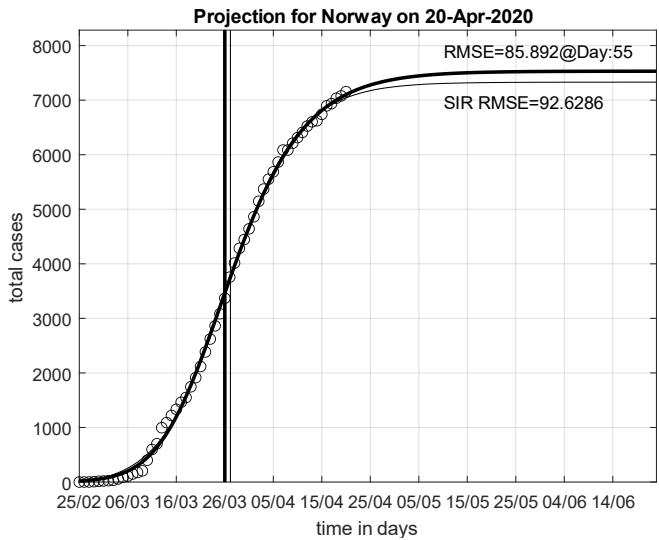
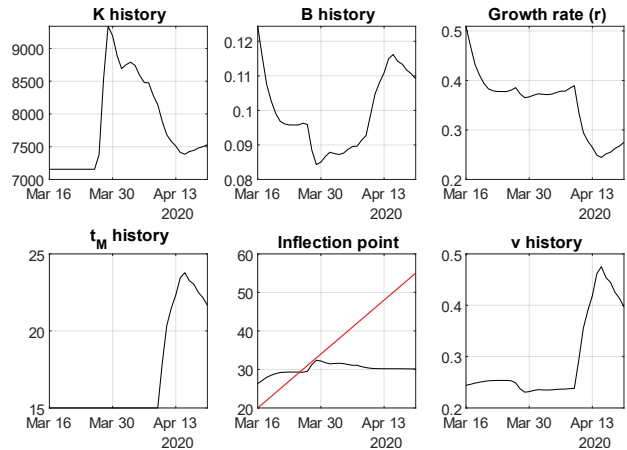

(a) Richards' versus SIR model projections with estimated turning points

(b) Richards' model parameters history.

**Figure 8.** Modelling results for Norway on 20.04.2020

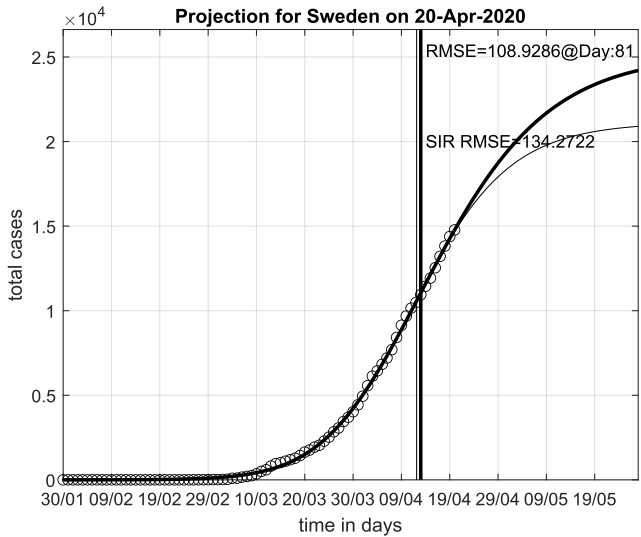
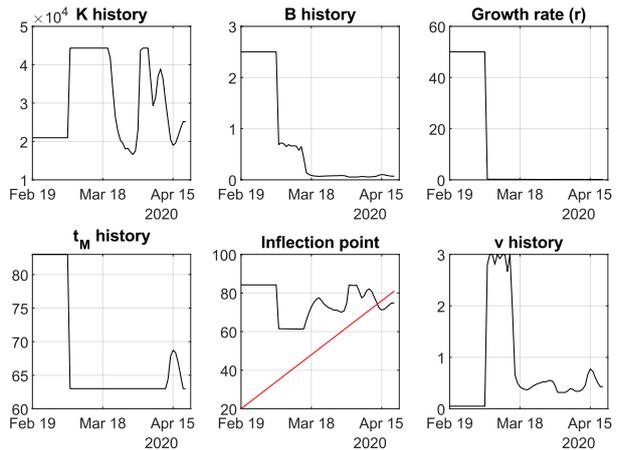

(a) Richards' versus SIR model projections with estimated turning points

(b) Richards' model parameters history.

**Figure 9.** Modelling results for Sweden on 20.04.2020

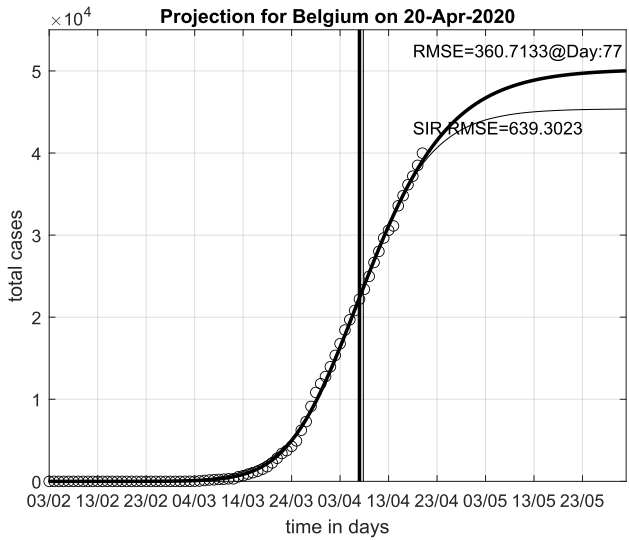
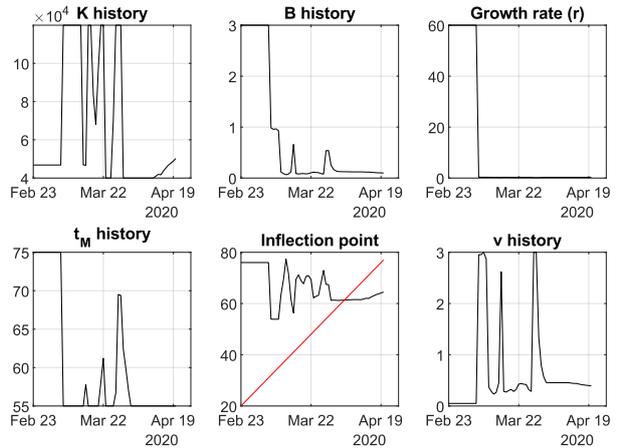

(a) Richards' versus SIR model projections with estimated turning points

(b) Richards' model parameters history.

**Figure 10.** Modelling results for Belgium on 20.04.2020

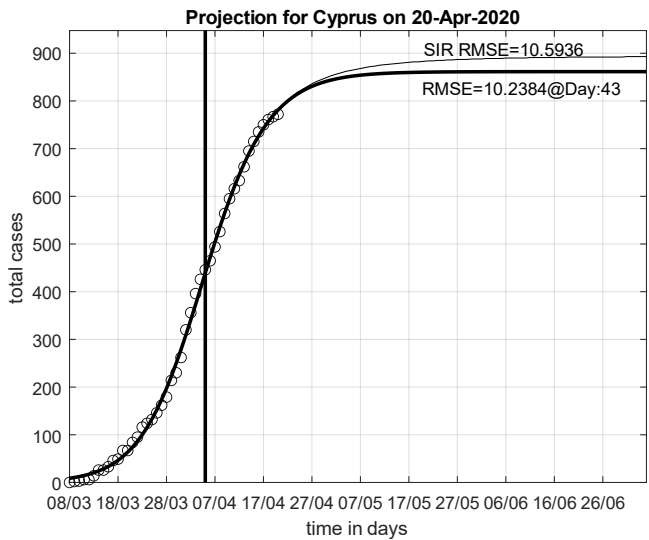
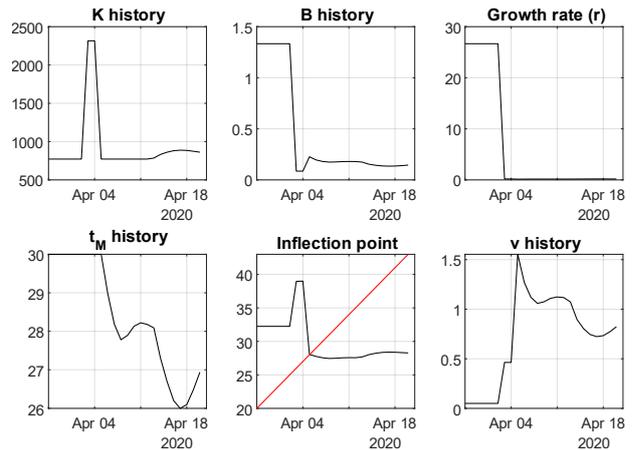

(a) Richards' versus SIR model projections with estimated turning points

(b) Richards' model parameters history.

**Figure 11.** Modelling results for Cyprus on 20.04.2020

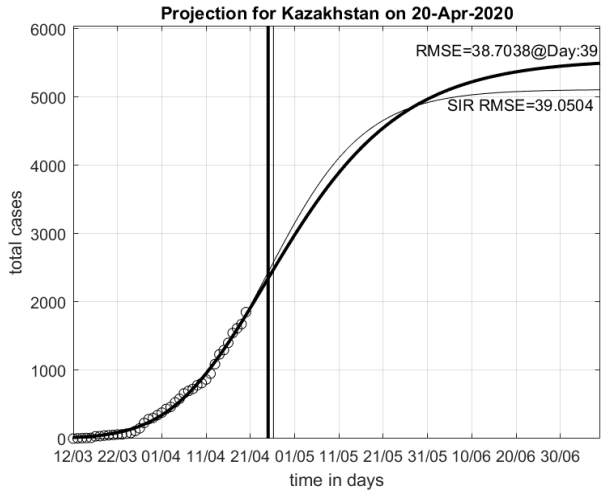
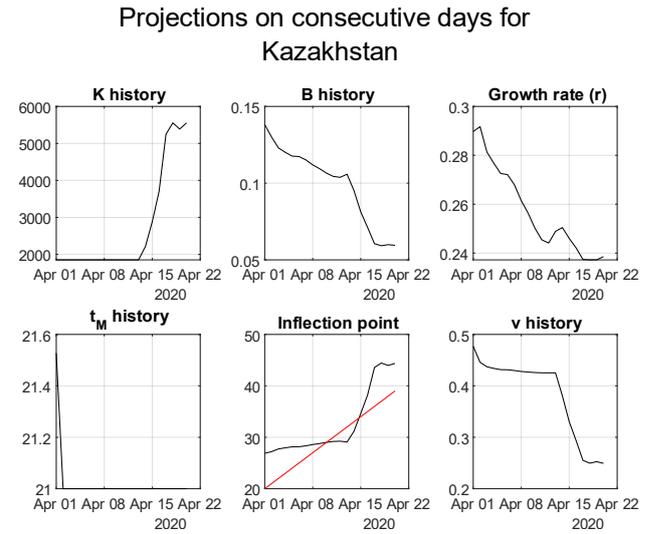

(a) Richards' versus SIR model projections with estimated turning points

(b) Richards' model parameters history.

**Figure 12.** Modelling results for Kazakhstan on 20.04.2020

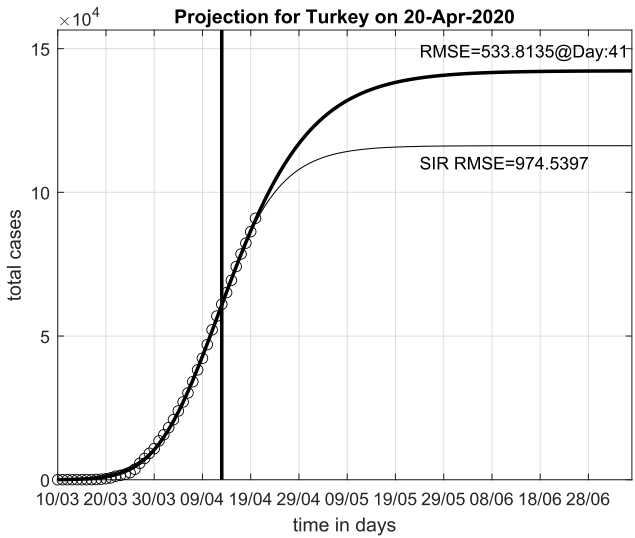
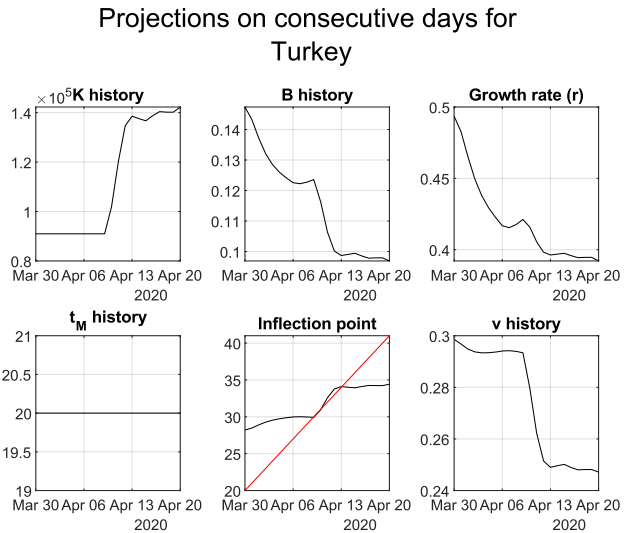

(a) Richards' versus SIR model projections with estimated turning points

(b) Richards' model parameters history.

**Figure 13.** Modelling results for Turkey on 20.04.2020

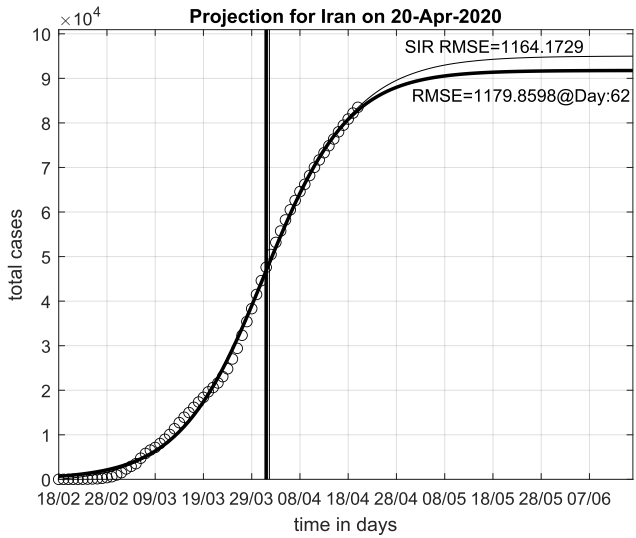

(a) Richards' versus SIR model projections with estimated turning points

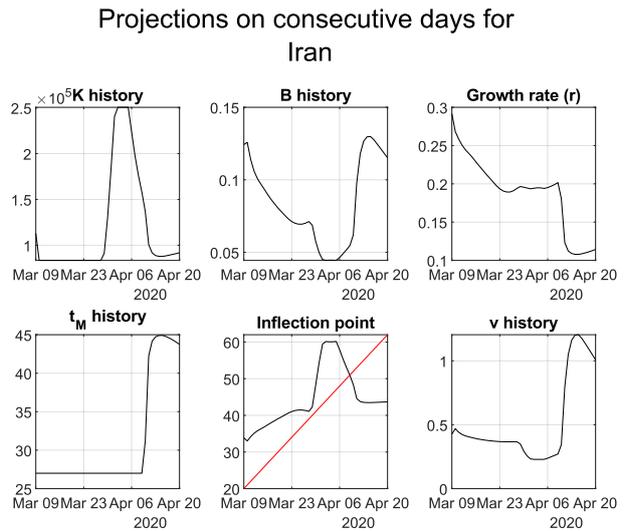

(b) Richards' model parameters history.

**Figure 14.** Modelling results for Iran on 20.04.2020

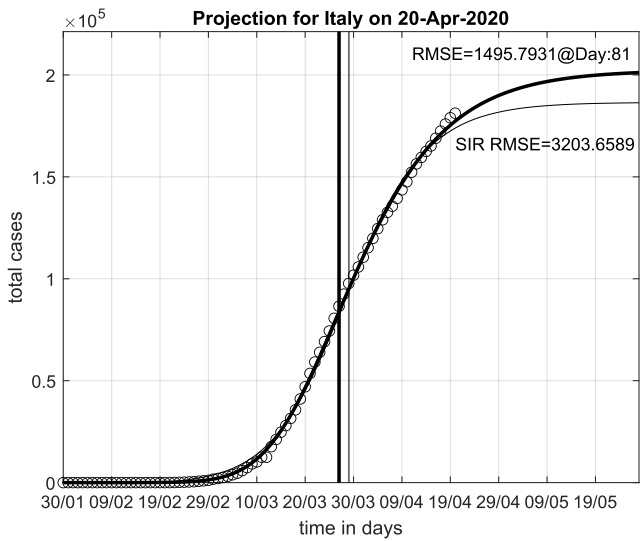

(a) Richards' versus SIR model projections with estimated turning points

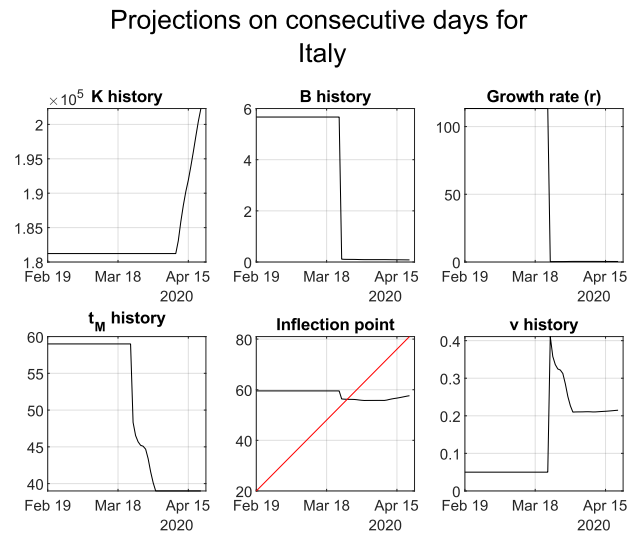

(b) Richards' model parameters history.

**Figure 15.** Modelling results for Italy on 20.04.2020

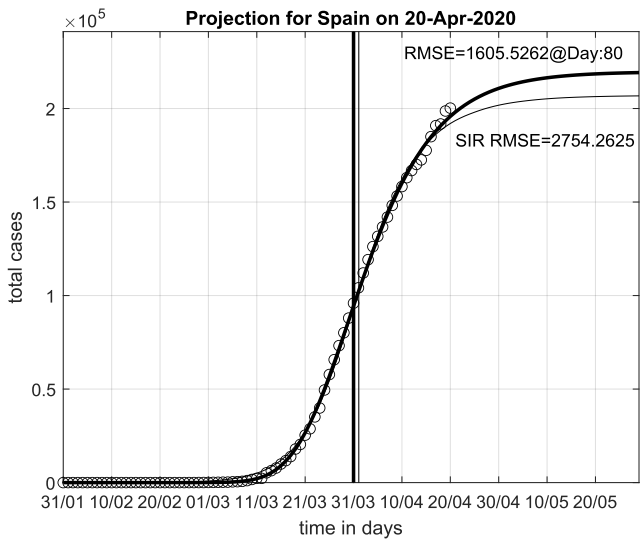
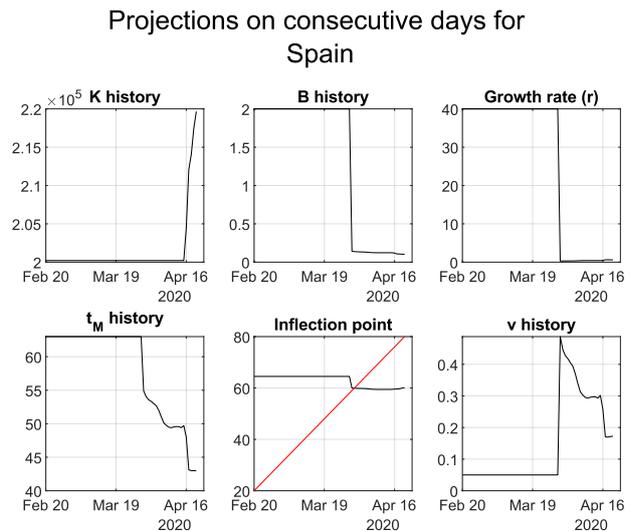

**(a)** Richards versus SIR model projections with estimated turning points

**(b)** Richards model parameters history.

**Figure 16.** Modelling results for Spain on 20.04.2020

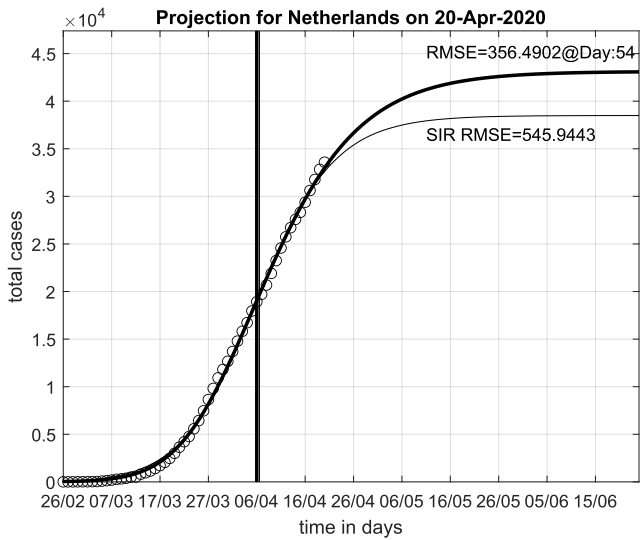
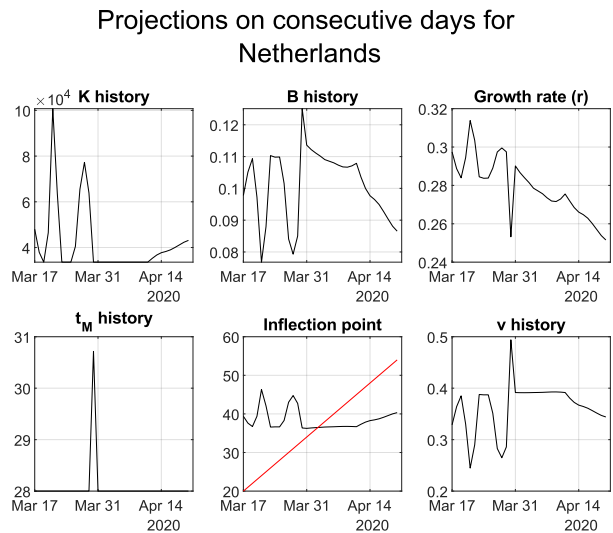

**(a)** Richards versus SIR model projections with estimated turning points

**(b)** Richards model parameters history.

**Figure 17.** Modelling results for Netherlands on 20.04.2020

## Discussion

### Testing and projections

Figures 4 and 5 demonstrate the evolution of deaths, confirmed cases and deaths-to-cases ratio for the selected countries. These graphs can be used to evaluate the outbreak severity and the effect of measures taken, including number of tests in each country. Particularly useful is the deaths to cases ratio in countries with high levels of testing, which includes symptomatic and asymptomatic patients. For these countries, the results can lead to a more accurate Case Fatality Ratio (CFR) calculation. The number of tests per confirmed case for selected countries (given availability of information), are shown in Table 1:

**Table 1. Tests per confirmed case for selected countries [17], [19], [20].**

| Country name | Tests / case | Date of reporting (2020) |
|---|---|---|
| Cyprus | 47.3 | April 21 |
| Greece | 24.1 | April 18 |
| Italy | 8 | April 21 |
| Turkey | 7.8 | April 21 |
| Sweden | 6.8 | April 19 |
| Netherlands | 5.4 | April 20 |
| USA | 5.3 | April 20 |
| Belgium | 4.5 | April 19 |
| Iran | 4.4 | April 21 |
| United Kingdom | 3.2 | April 21 |

Therefore, the deaths-to-cases ratio can provide a relatively accurate depiction of current CFR in those countries with high number of tests per case. At the time of writing, this is valid for Greece and Cyprus, and at a lesser extent, for Italy, Sweden, and Turkey. Furthermore, in the countries with high number of tests per case, projections are more representative of the actual number of cases and may be less impacted by "hidden" cases of asymptomatic people. In Cyprus, which has the highest number of tests per case in our table, the current CFR estimate equals 1.6% but still, additional testing might be required for identifying asymptomatic people and establishing a more accurate figure; follow-up studies will consider the updated information. Conclusions regarding the accuracy of reporting and testing can also be drawn from Figure 4. For example, in Iran, the first death due to COVID-19 appears on the same day as the first case, which would not make sense since an average number of 18 days is reported as the time from diagnosis to death for these patients [16]. In Turkey and Netherlands this delay period is around a week, which might signify a late confirmation of the first cases, while for the remaining of the examined countries, a period of two weeks or more is recorded, which is more reasonable given our current understanding of the disease. Another important consideration is the slope of the death to cases ratio. Constant deaths-to-cases ratios (zero or small slope values) could signify effective testing and accurate reporting and/or efficient handling of the disease by the corresponding healthcare systems. On the other hand, in Sweden and Belgium, as an example, the deaths-to-cases ratio keeps increasing [17], which could be due to reporting inconsistencies and/or exceeding healthcare system's capacity. Specifically, on April 20$^{th}$ the healthcare systems of both countries[4] had considerably less available ICU beds than required [18]. Moreover, and especially for the case of Sweden, if we look at the daily cases reporting diagram in Figure 18, it is evident that a weekly consistent pattern with less reported cases during the weekends appears. This is possibly an indication[5] for co-existence of inconsistent reporting that affects the quality of the timeseries we are using.

Therefore, given the above discussion, graphs appearing in Figures 4 and 5 along with the testing levels in Table 1 can help place the projections in the right perspective. For example, if a country is underreporting cases it is possible that at some point the projection will "jump" to a new level after a surge in the number tests and the discovery of clusters of asymptomatic patients. That is why in follow-up cases it is important to quantify the testing efficiency and adjust the data accordingly.

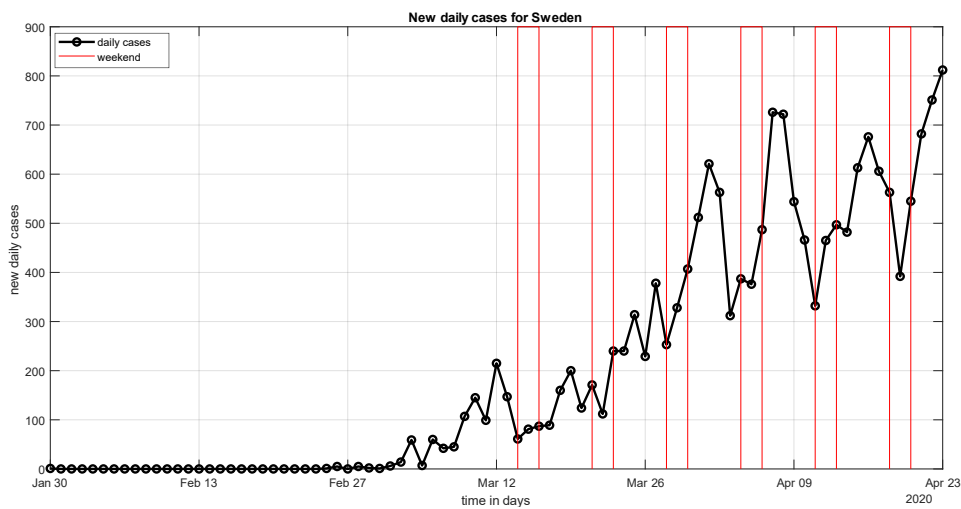

**Figure 18:** Oscillatory behavior in new daily cases for Sweden

**SIR model versus Richards model**

The Richards model projections are generally in alignment with the ones generated by the SIR model with a consistently better fit (lower Root-Mean-Square-Error - RMSE) reported for the former. The differences between the projections by the two models, especially when it comes to the capacity $K$ and the timing of the inflection point, $t_i$, fluctuate over the history of projections, but consistently remain sufficiently close. This indicates that both models, and for the countries examined, have a similar predictive capacity in assessing the current outbreak severity, forecasting the expected number of new cases in the coming days, and estimating the time of the turning point. Of course, predicting the future

---

[4] see https://covid19.healthdata.org/sweden and https://covid19.healthdata.org/sweden
[5] The same oscillatory behavior appears also in the daily deaths diagram; see https://www.worldometers.info/coronavirus/country/sweden/

"*is a risky but tantalizing endeavor*" [7], and that's why it is widely accepted that projections and even more, forecasts, should only be employed for assessing the current state of the outbreak, implementing containing measures more effectively and drafting a general strategy with reference to projected future outcomes.

**History of parameters**

The generated history of parameter values, i.e., the model's meta-projections, can provide significant insights regarding the confidence in our current projections and the overall development of the epidemic. For example, when examining the parameters' trajectories for the herein modeled countries, we can easily observe that for the cases of USA, Italy, Spain Turkey and Kazakhstan there is a clear time period, close to our current observation time, with a constantly increasing projected number of the total number of cases ($K$) at the end of the employed 120-days period; although Turkey and Kazakhstan seem to be currently stabilizing this projected number. For Greece and Iran, on the other hand, a current significant decrease in $K$ values has been recorded. Given each country's circumstances, the current state of projected $K$ can mean different things. Take Cyprus as an example that has a high number of tests per case and has potentially reached a point at which the majority of COVID-19 cases are identified. At the same time, its day-to-day projected $K$ appears constant, and this can signify stable containment of the virus transmission. On the other hand, USA, with a more or less constant testing frequency during the last two weeks [17], exhibits increasing projected $K$ values, and this could potentially mean that the transmission is still at the growth phase and not under control. At the same time, the agreement in tendencies with the SIR model observed in Figure 6 for USA excludes the possibility that the increase is due to model's sensitivity for the specific date. In a fully developed and contained phase, all parameters stabilize and converge to constant values as can be seen for the case of China in Figure 19.

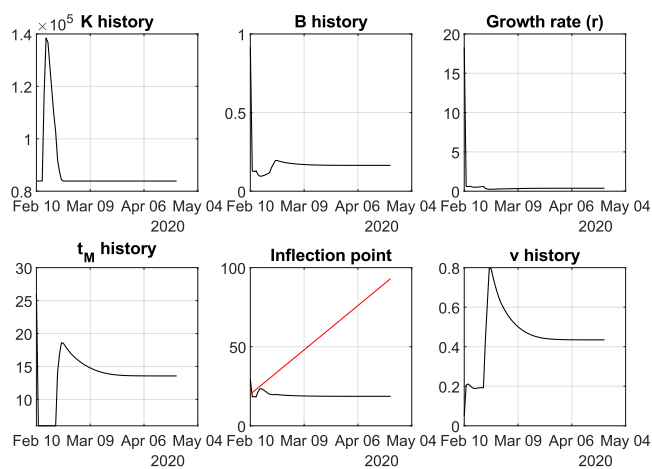

**Figure 19.** Converged history of parameters for China.

Following the inflection point's history is also a very significant element in these meta-projections. As it can be seen in all examined cases, clearly passing the turning point (leaving it in the past) is accompanied by a clear drop in the growth rates. Some countries, like Kazakhstan seemed temporarily to have passed the turning point but the epidemic underwent a transition phase and now it is clear that the turning point is still in the future.

In addition to evaluating each country's state of the pandemic individually, some similarities can be observed in the parameters' history profiles between countries. For example, Greece, Norway and Iran, and at a lesser extent Cyprus share a similar profile; see Figure 20. In all of these countries, the turning point resides in the past and a considerable drop in the total number of cases occurred after passing the turning point. Currently, all of them seem to be on a stable development with an expected convergence of the model parameters in the following days.

US, Italy and Spain (see Figure 21) seem to have also passed the turning point but the projection for the total number of cases continues to increase, in a more or less, steady rate with no indicated epidemiological transitions. The situation in these countries seems to be developing steadily and additional data and successive estimates will be needed to get convergence and safely predict the final numbers. Sweden and Belgium, on the other hand, exhibit an erratic behavior in the corresponding parameter histories as can be seen in Figure 22. One may argue that this is a modelling failure, but we also need to consider some special circumstances for these two cases. As we have previously mentioned, both countries are facing serious problems with the capacity of their healthcare systems, although this does not seem to be the primary reason for this behavior since otherwise Italy and Spain would share similar profiles with them. Epidemiological transitions and local destabilizing outbreaks could also explain the situation since both countries are currently close to their respective turning points, which at the same time is a hard point for accurate projections. Additionally, for the case of Sweden, there is an indication of irregular case reporting which could affect the quality of the data and consequently the quality of our projections. Moreover, Sweden's decision to avoid some of the strict measures other countries are taking, may render the country more vulnerable to local destabilizing outbreaks. Belgium is maybe the worst hit country[6] in the world on the basis of deaths per million of inhabitants[7] and one can easily imagine that retaining control of the situation would be extremely difficult under such conditions which could also partially explain this erratic behavior in successive projections for Belgium. Finally, Turkey and Kazakhstan seem to be at an early stage of the epidemic development and few things can be said with certainty; see Figure 23. Nevertheless, the parameters' trajectories resemble the trajectories of the countries in Figure 20 up to the point where $K$ peaks, and therefore, certain outbreak dynamics can be anticipated or planned-for with appropriate containment measures.

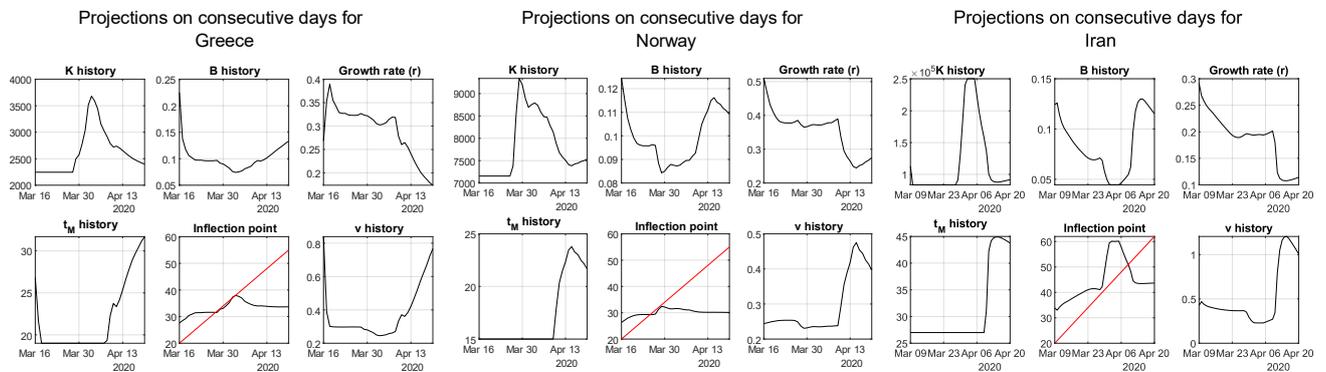

**Figure 20.** Comparison of history of parameters for Greece, Norway, and Iran.

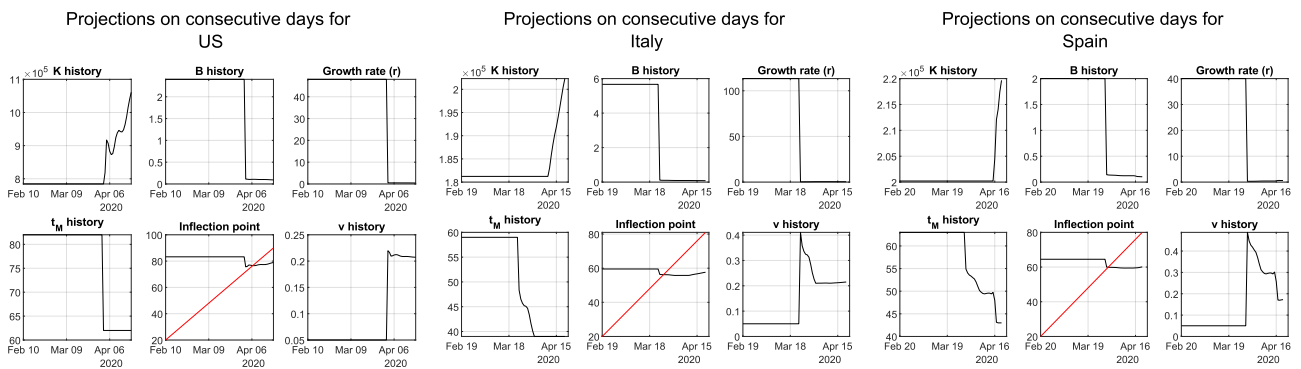

**Figure 21.** Comparison of history of parameters for US, Italy and Spain.

---

[6] Next to San Marino
[7] More than 580 deaths per million of inhabitants [2].

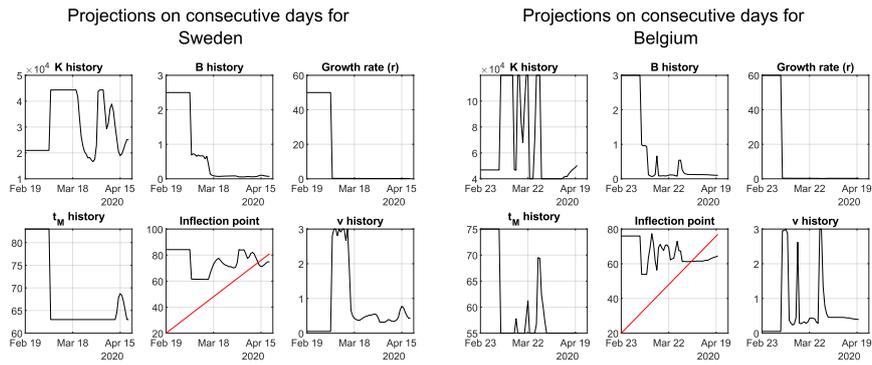

**Figure 22.** Comparison of history of parameters for Sweden and Belgium.

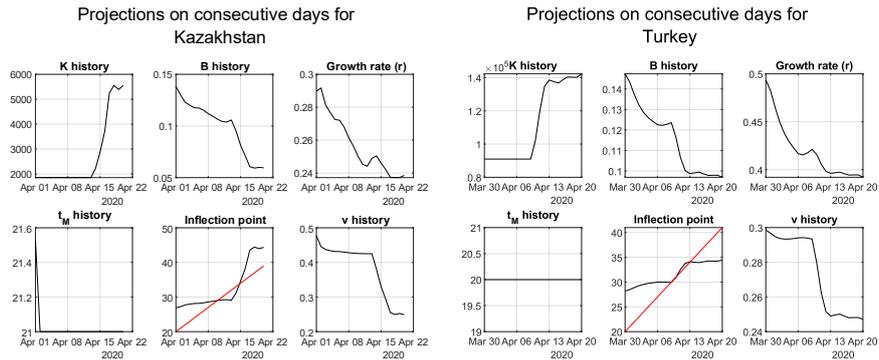

**Figure 23.** Comparison of history of parameters between Kazakhstan and Turkey.

# Conclusions

In this work we have demonstrated that Richards' curve, although simple, has sufficient predictive power when used to model disease outbreaks. Based on our comparisons with one of the most common and simple tools for mathematically modelling infectious diseases (SIR), we have provided evidence that Richards curve has an equivalent modelling capacity while being equivalently simple to SIR, if not simpler. Moreover, we have demonstrated that the use of meta-projections, i.e., histories and the corresponding trends of Richards' model parameters from successive projections, can provide researchers and professionals with insights on assessing the appropriateness of the employed models, effortlessly identify different phases of the disease and in effect decrease the uncertainty in their estimations of an epidemic's severity and development. A number of different countries with successive projections regarding COVID-19 development till 20.04.2020 have been included and discussed. This discussion naturally brought up the issue of sufficient and reliable data which is an absolute pre-requisite to attain projections which could be potentially useful and meaningful.